\documentclass{appolb}

\usepackage{epsfig, rotating}

\begin{document}

\title{$\Lambda$-hypernuclear production in ($K^-_{\rm stop}, \pi$) reactions%
\thanks{This work was supported by the GAUK grant No. 91509 and the GACR grant No. 202/09/1441}%
}

\author{Vojt\v{e}ch Krej\v{c}i\v{r}\'{i}k
\address{Nuclear Physics Institute, \v{R}e\v{z}, Czech Republic\\ Faculty of Mathematics and Physics, Charles University, Prague, Czech Republic}
\and
Ale\v{s} Ciepl\'{y}
\address{Nuclear Physics Institute, \v{R}e\v{z}, Czech Republic}
}

\maketitle

\begin{abstract}
We report on calculation
of the $\Lambda$~-~hypernuclear production induced by the stopped $K^-$. The 
calculation was performed within the framework of the distorted
wave impulse approximation and employed chirally motivated model for the microscopic description of
the elementary $K^-$-nucleon process.
The sensitivity of the calculation was tested with
various wave functions of both the kaon in the initial state and
the pion in the final state. Our results
are closer to the experimental values then the 
results of previous calculations.
\end{abstract}


\section{Introduction}

We studied the $\Lambda$-hypernuclear production induced by the stopped kaon,
$(K^-_{\rm stopped},\pi)$. We believe that 
an analysis of this process can provide additional information on a depth of the $K^-$-nucleus potential,
whether it is deep or shallow.
Previous calculations \cite{GalKlieb, MatsuyamaYazaki, CieplyFriedman} did not give satisfactory predictions, the capture rates
were at least three times smaller then the experimental values.
The novelty in our approach is mainly a microscopic description of the elementary process using a chirally motivated
model.
Moreover, we studied the sensitivity of the calculation to various input wave functions.

\section{Formalism}

We used a distorted wave impulse approximation (DWIA) as described in detail by
Gal and Klieb~\cite{GalKlieb}. The T-matrix is written in a form
\begin{equation}
T_{\rm if}({\bf q_{\rm f}}) = t_{\rm if}({\bf q_{\rm f}}) \,\, \int  {\rm d^3} {\bf r} \,\, \chi_{q_{\rm f}}^*({\bf r}) \,\, \rho_{\rm if}({\bf r}) \,\, \Psi_{NLM}({\bf r}) \,\, .
\label{TMatElement}
\end{equation}
Here, $t_{\rm if}({\bf q_f})$ denotes the t-matrix of the elementary process,
$\Psi_{NLM}({\bf r})$ and $\chi_{q_{\rm f}}({\bf r})$ are the $K^-$ and $\pi$ wave functions, 
${\bf q_{\rm f}}$ is the pion momentum, and
$\rho_{\rm if}$ stands for the nucleus to hypernucleus transition density matrix.

The capture rate per one stopped kaon $R_{\rm if}$ is 
defined as a ratio of the capture rate for a specific process to the total capture rate.
After some manipulations (see \cite{GalKlieb}), the final formula for $R_{\rm if}$ can be expressed as
a product of three terms:
\begin{equation}
R_{\rm if} = { { q_{\rm f} \omega_{\rm f} }\over{ \overline{q}_{\rm f} \overline{\omega}_{\rm f} } } \cdot R(K^- N\rightarrow \pi Y) \cdot R_{\rm if}/Y .
\label{CRperKaon}
\end{equation}
The first term  is a kinematical factor,
the second 
represents the branching ratio for the elementary process and
the third, which we call the capture rate per hyperon, reads 
\begin{equation}
R_{\rm if}/Y =\frac{\int \frac{ d\Omega_{q_{\rm f}} }{ 4 \pi } \left< \left| \int  {\rm d^3} r \,\, {\chi_{q_{\rm f}}^{(-)} }^*({\bf r}) \,\, \rho_{\rm if}({\bf r}) \,\, \Psi_{NLM}({\bf r}) \right|^2 \right>   }{\widetilde{\rho}_N} ,
\label{rateperhyperon1}
\end{equation}
where
\begin{equation}
\widetilde{\rho}_N = \int {\rm d^3} {\bf r} \,\, \rho_N(r) \, \rho_{K^-}(r)\, \rho_{\pi}(r) 
\label{efektivnihustota}
\end{equation}
is called the effective nucleon density. Following Gal and Klieb \cite{GalKlieb} one should replace the pion distribution by
a simple plane wave ($\rho_{\pi}~=~1$) in Eq. (\ref{efektivnihustota}) in order to account for
all possible final states contributing to the total capture rate.
We also looked at the effect of keeping the pion distortion in Eq. (\ref{efektivnihustota}). It leads to a substantial increase of
the calculated capture rates and to a better agreement with experimental data. Although we find this feature interesting,
such modification is not consistent with the DWIA approach and therefore cannot be taken seriously.

The factor $R_{\rm if}/Y$ is evaluated analytically using 
the spherical coordinates and the partial wave expansion. The final form reads
\begin{equation}
R_{n_Nl_N \rightarrow n_Yl_Y}/Y = \frac{N(j_N) \sum_k (2k+1)(l_N 0 k 0 | l_Y 0) N^{(k)}_{\gamma_Y\gamma_N} }{\int {\rm d}r \,\, \rho_N(r) \, |R_{NL}(r)|^2} \,,
\label{CRperHYP}
\end{equation}
where
$$
N^{(k)}_{\gamma_Y\gamma_N} = \sum_{l} (L 0 \, k 0 | l 0)^2 | I^l_{\gamma_Y\gamma_N} |^2 \,\,,
$$
$N(j_N)$ is the number of nucleons in the shell $j_N$, and $I^l_{\gamma_Y\gamma_N}$ stands for the overlap of
the radial parts of the wave functions of $K^-$, $N$, $\pi$, and $Y$.

\section{Inputs}

In this chapter, we present
the elementary branching ratios and 
potentials used to determine the wave functions of $K^-$, $\pi$, $N$ and $Y$.

\subsection{Elementary branching ratios}

In order to describe the elementary $K^-$-nucleon process,
we adopted the effective potential model based on chiral symmetry. The details of
the approach can be found in Refs. \cite{KaiserSiegelWeise, WaasKaiserWeise, CieplySmejkal, CieplySmejkalPrep}. 
In the calculation, we took account of the Pauli blocking \cite{WaasKaiserWeise} and the $K^-$-selfenergy \cite{CieplyFriedman}.
The pertinent elementary branching ratios evaluated at the $K^-N$ threshold 
are $R(K^-n~\rightarrow~\pi^-~\Lambda)~=~10.72$, and $R(K^-p~\rightarrow~\pi^0~\Lambda)~=~5.36 $.

\subsection{Wave functions}

To calculate the $K^-$-atomic wave function, we used an optical potential, which describes the strong interaction, in
addition to the electromagnetic interaction (including finite size charge distribution and vacuum polarization effects). 
The strong interaction $K^-$-nucleus potential is taken in the form devised in Ref. \cite{Batty},
\begin{equation}
V^{K}_{opt}(r) = - \frac{4 \pi}{2 \mu} \left(1+ \frac{\mu}{M_N} \right) \left[ b + B \left(  \frac{\rho(r)}{\rho(0)}  \right)^{\nu} \, \right] \rho(r).
\label{optpot}
\end{equation}
We used three different parameter sets, which are specified in Table~\ref{TabKaonOptPot}. 
The choice $[K_{\chi}]$ represents the chiral model \cite{CieplySmejkal},
$[K_{\rm eff}]$ and $[K_{\rm DD}]$ denote phenomenological potentials taken from
Ref. \cite{Batty}. For a reference, we also show the respective potential depths in the last column of the table.
Moreover, we also performed the calculation with a pure electromagnetic potential ([$K_{\rm EM}$]) 
to check the impact of the strong interaction.

\setcounter{table}{0}

\begin{table}[h!]
\begin{center}

\caption{Parameters of the kaonic optical potential.}
\label{TabKaonOptPot}

\begin{tabular}{| l |c |c |c |c|}
\hline
set  &  $b~[{\rm fm}]$ &  $B~[{\rm fm}]$  &  $\nu$ & $V^{K}_{opt}(\rho=\rho_0$)\, [MeV] \\\hline
$[K_{\chi}]$ & $0.38+0.48{\rm i}$ & $0$ & $0$ &   50  \\\hline
$[K_{\rm eff}]$ & $0.63+0.89{\rm i}$ & $0$ & $0$ &  80 \\\hline
$[K_{\rm DD}]$     &  $-0.15 + 0.62{\rm i}$ & $1.65-0.06 {\rm i}$  &  $0.23$ &  190 \\\hline
\end{tabular}

\end{center}
\vspace*{-1em}
\end{table}

The baryon wave functions were obtained using Wood-Saxon potential with parameters fixed to reproduce single particle binding energies.
The pion-nucleus optical potential is taken in the standard form \cite{EricsonEricson}.
We performed our calculations for a free pion ($\pi_0$) and for two different parameter sets,
 $(\pi_{\rm b})$ \cite{pionoptpot1} and $(\pi_{\rm c})$ \cite{pionoptpot2}, which describe the low energy pion scattering data.

\section{Results and discussion}

In this chapter, we present our results and 
discuss the sensitivity of the calculated capture rates to the choice of the $K^-$-nucleus and $\pi$-nucleus
potential.
We focus on the production of $\Lambda$~-~hypernuclei
($^{12}_{\Lambda}{\rm C}$, $^{12}_{\Lambda}{\rm B}$, $^{16}_{\Lambda}{\rm O}$, $^{16}_{\Lambda}{\rm N}$) and 
take into account the hyperon formation in both the $1S$ and the $1P$ states.

\begin{figure}[h!]
\begin{center}

\includegraphics[width=7cm]{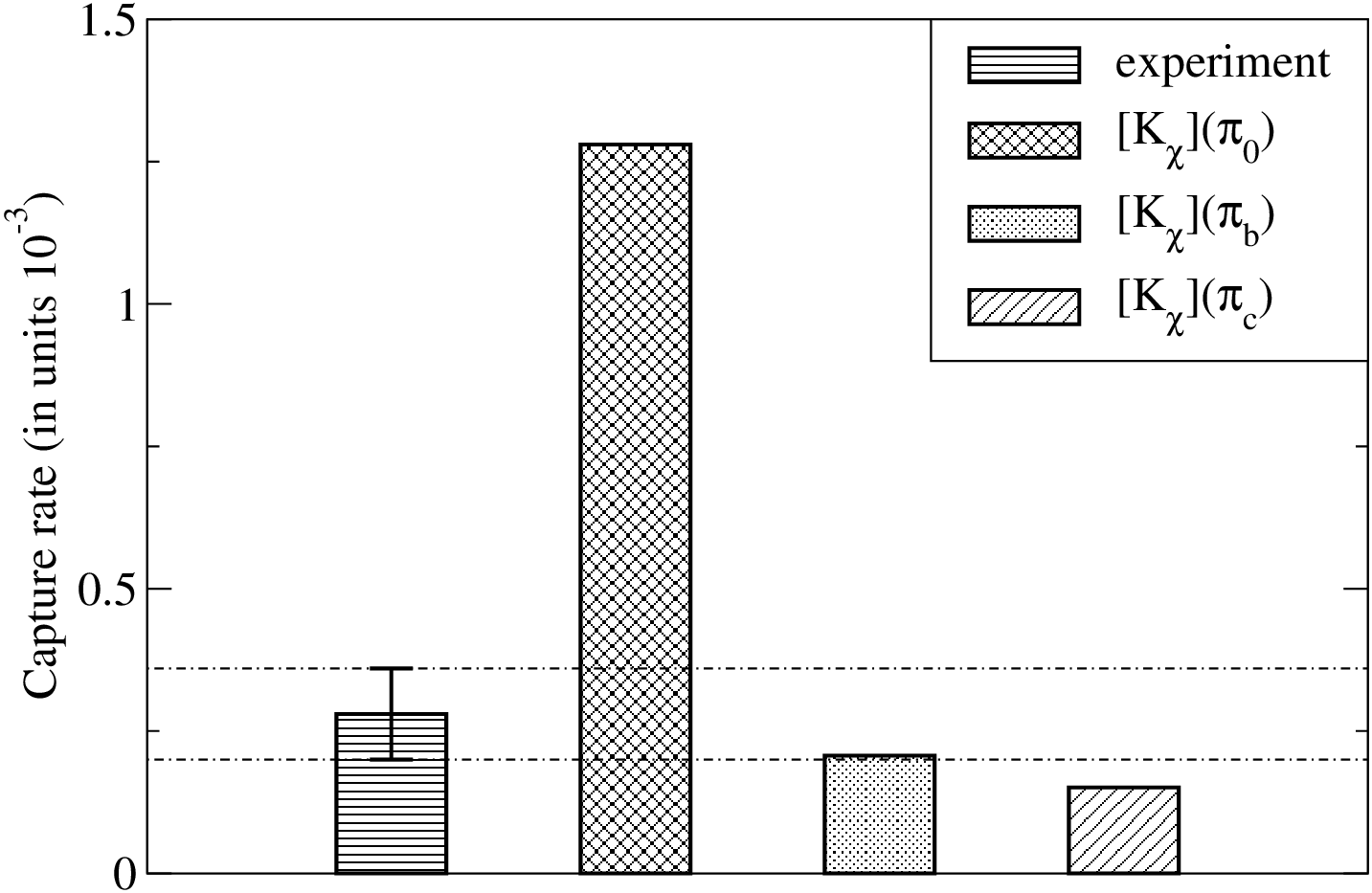} 

\caption{The production of $^{12}_{\Lambda}{\rm B}$ in the $1S_{\Lambda}$ state.}
\label{FigPionSens}

\end{center}
\vspace*{-1em}
\end{figure}

First, we look at the effect of various pion wave functions.
Our results are shown in Fig.\ref{FigPionSens}.
One can see that the inclusion of the pion distortion, no matter whether ($\pi_{\rm b}$) or ($\pi_{\rm c}$),
leads to a substantial
decrease of the capture rate $R_{\rm if}$ (up to one order).
Apparently, the final state interaction plays an important role.
On the other hand, it looks that the computed rates are not very sensitive to the choice of pion-nucleus 
optical potential.

\begin{figure}[h!]
\begin{center}

\includegraphics[width=7cm]{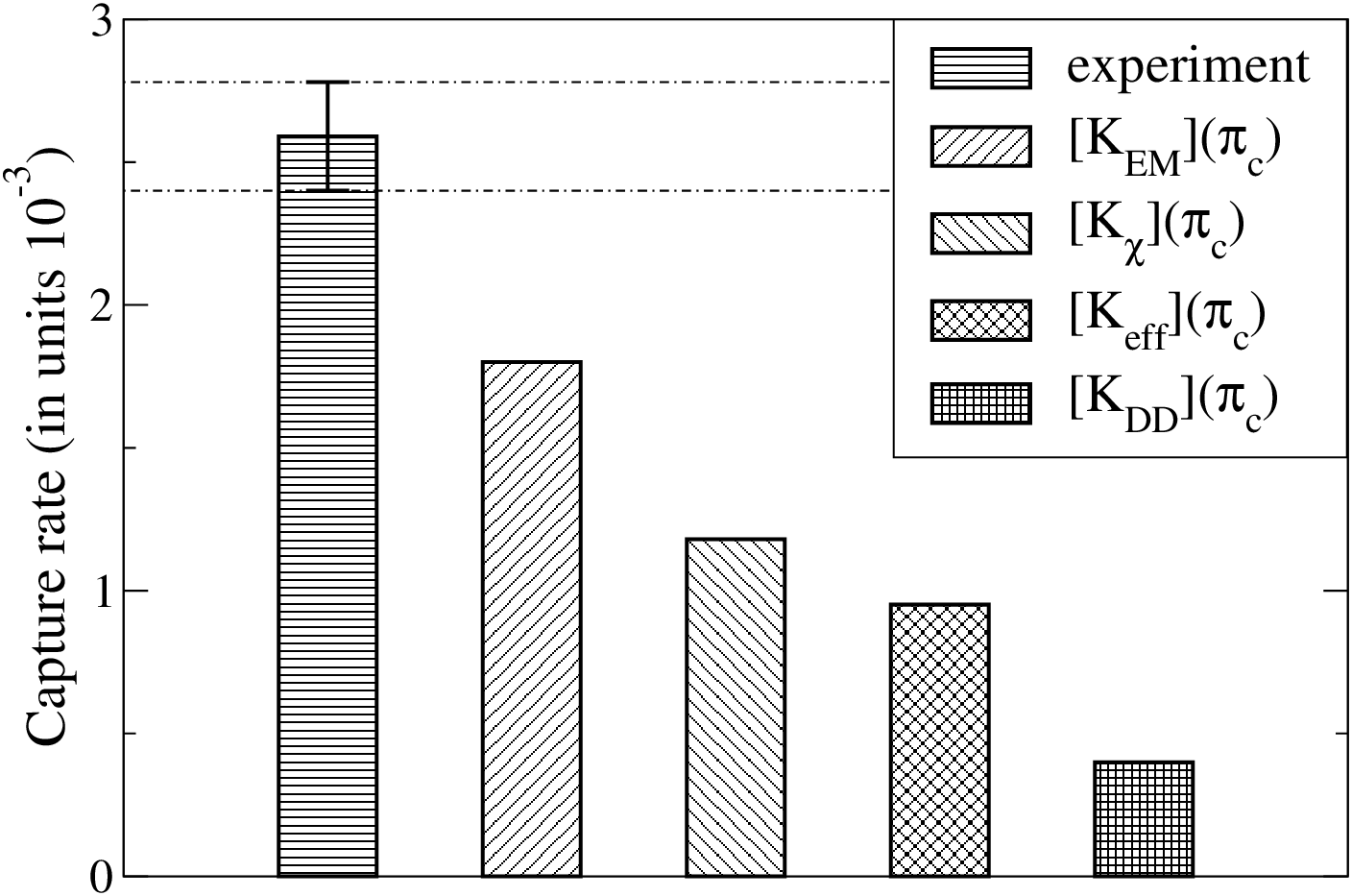}

\caption{The production of $^{12}_{\Lambda}{\rm C}$ in the $1P_{\Lambda}$ state.}
\label{FigKaonSens}

\end{center}
\vspace*{-1em}
\end{figure}

The sensitivity of the capture rates to the choice of $K^-$ wave functions is demonstrated in Fig.\ref{FigKaonSens}.
It appears that the capture rate is a decreasing function of the $K^-$~-~nucleus potential depth.

To determine
the best combination of potentials involved in our work, 
we compare our results with the available experimental data for 
the production of $^{12}_{\Lambda}{\rm C}$ \cite{Agnello}, $^{12}_{\Lambda}{\rm B}$ \cite{Ahmed},
and $^{16}_{\Lambda}{\rm O}$ \cite{Tamura}.
In the comparison, we include six production rates to the 
 $1S_{\Lambda}$ and $1P_{\Lambda}$ states
and four ratios between them.
The resulting $\chi^2$ per data point are shown in Table~\ref{TabComp}.
The best value is achieved for the combination of the $[K_{\chi}]$ and $(\pi_b)$ potentials. It is interesting that
comparable (if not even better) results are obtained with the kaon wave function generated by a purely electromagnetic 
interaction (the choice [$K_{\rm EM}$]).

\begin{table}[h!]
\begin{center}

\caption{The comparison of various combinations of potentials.}

	\begin{tabular}{  |c|cccccccc| }
	
	\hline
	$\chi^2 \,/\, N$&&   [$K_{\rm EM}$]	&&	[$K_{\chi}$]  	&&	[$K_{\rm eff}$]	&&	[$K_{\rm DD}$] 	 \\\hline
	($\pi_0$)	&&	206.7		&&	219.9		&&		166.0	&&	79.3	\\
	($\pi_{\rm b}$)	&&	7.3		&&	{\bf 7.7}	&&		11.7	&&	20.0	\\
	($\pi_{\rm c}$)	&&	7.9		&&	10.0		&&		14.2	&&	31.3	\\\hline
	\end{tabular}

\label{TabComp}

\end{center}
\vspace*{-1em}
\end{table}

\begin{figure}[h!]
\begin{center}

\includegraphics[width=5.6cm]{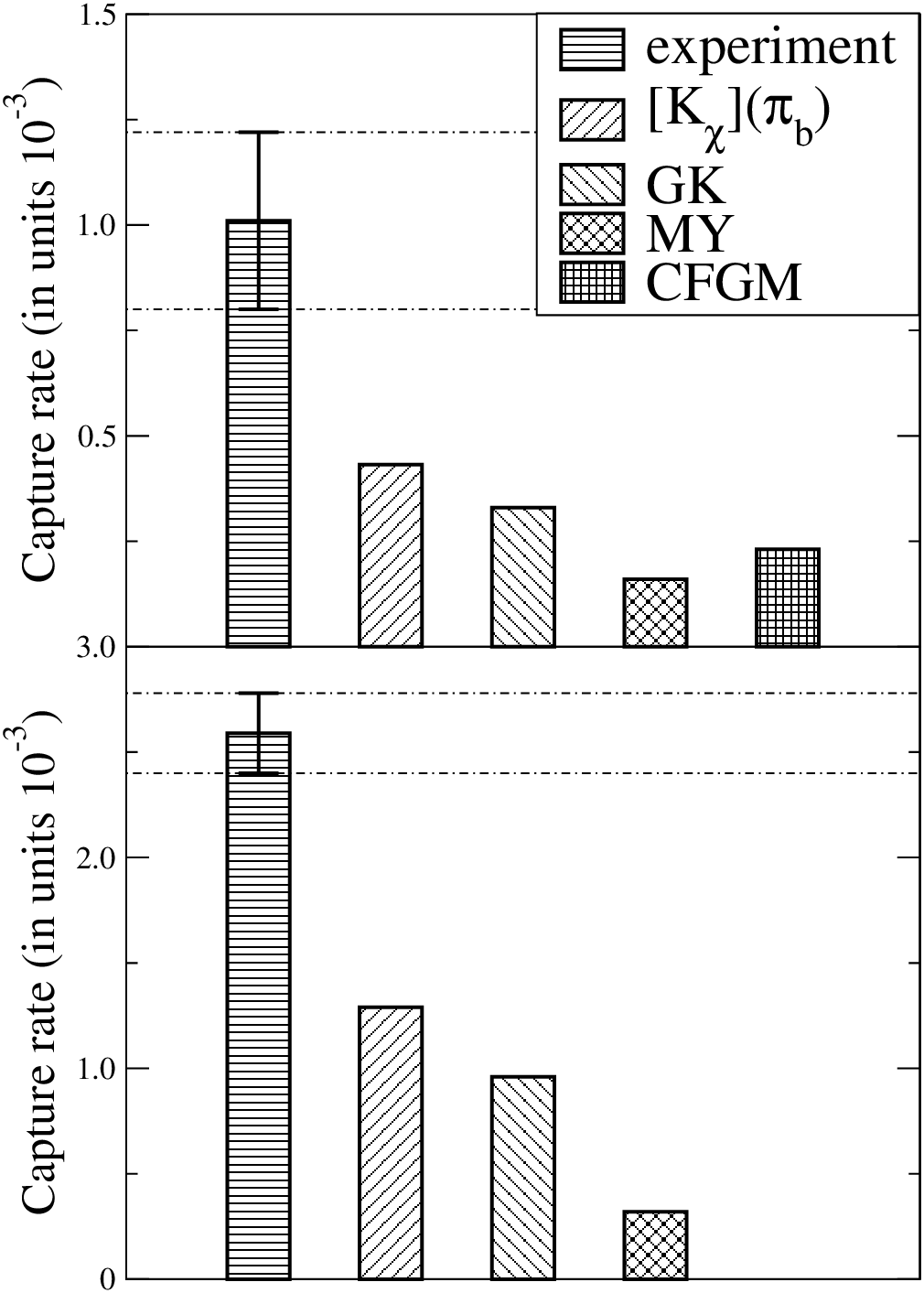} 
\hskip 2em
\includegraphics[width=5.6cm]{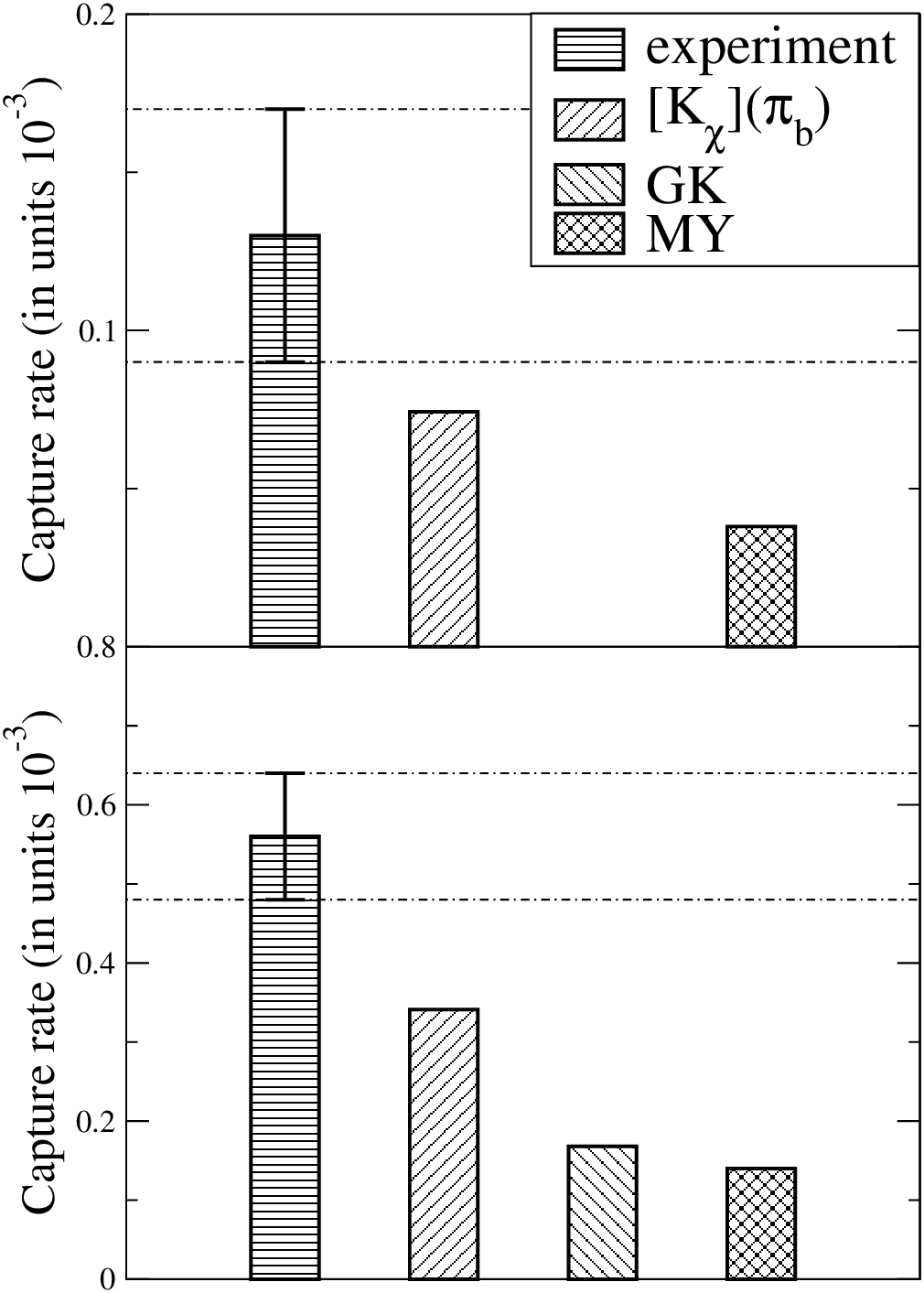}

\caption{The production of $^{12}_{\Lambda}{\rm C}$ (left) and $^{16}_{\Lambda}{\rm O}$ (right)
in the $1S_{\Lambda}$ (top) state 
and in the $1P_{\Lambda}$ state (bottom).}
\label{FigResCO}

\end{center}
\vspace*{-1em}
\end{figure}

The comparison of our results with experimental data \cite{Agnello, Ahmed, Tamura} 
and with the previous calculations is given in Fig.\ref{FigResCO}.
The predictions made by Gal and Klieb~\cite{GalKlieb} are labeled GL,
by Matsuyama and Yazaki~\cite{MatsuyamaYazaki} MY, and
by Cieply et~al.~\cite{CieplyFriedman} CFGM.
Apparently, our results are closer to the experimental data than the results of previous works.

\section{Conclusion}

We performed the calculation of the $\Lambda$~-~hypernuclear production making use of a microscopic model for the description of
the elementary process. We showed that the capture rate is
a decreasing function of the kaon-nucleus potential depth and that our model based on the DWIA is very 
sensitive to pion distortion in the final state.

Our results are closer to the experimental data than the results of previous calculations.
It looks that the shallow potential [$K_{\chi}$] is the best in the description of the hypernuclear production.
Unfortunately, the ambiguities in the input wave functions and in other factors involved in the theory make this statement
very weak and do not allow us
to decide convincingly
which potential (deep [$K_{\rm DD}$] or shallow [$K_{\chi}$]) is better in general.

\end{document}